\documentstyle{article}
\begin{document}

\def\frak{\cal }
\def\Bbb{\bf }

\title{``Haunted'' quantum contextuality}
\author{K. Svozil\\
 {\small Institut f\"ur Theoretische Physik,}
  {\small University of Technology Vienna }     \\
  {\small Wiedner Hauptstra\ss e 8-10/136,}
  {\small A-1040 Vienna, Austria   }            \\
  {\small e-mail: svozil@tph.tuwien.ac.at}\\
  {\small www: http://tph.tuwien.ac.at/$\widetilde{\;\;}\,$svozil}}
\date{ }
\maketitle

\begin{flushright}
{\scriptsize http://tph.tuwien.ac.at/$\widetilde{\;\;}\,$svozil/publ/context.$\{$ps,tex$\}$}
\end{flushright}

\begin{abstract}
Two entangled particles in threedimensional
Hilbert space (per particle) are considered in an EPR-type arrangement. On each side
the  Kochen-Specker observables $\{J_1^2,J_2^2,J_3^2\}$ and $\{\bar J_1^2,
\bar J_2^2,J_3^2\}$ with $[J_1^2,\bar J_1^2]\neq
0$ are measured.
The outcomes of measurements of  $J_3^2$ (via $J_1^2,J_2^2$) and $J_3^2$
(via $\bar J_1^2,\bar J_2^2$) are compared.
We investigate the possibility that,
although formally $J_3^2$ is associated with the same projection
operator,
a strong form of quantum contextuality states that an outcome
depends on the complete disposition of the measurement apparatus,
in particular whether $J_1^2$  or $\bar J_1^2$ is measured alongside.
It is argued that in this case it is impossible to measure contextuality directly,
a necessary condition being a non-operational counterfactuality of the argument.
\end{abstract}


Besides complementarity, contextuality
\cite{bell-66,hey-red,redhead,peres,mermin-93}
is another, more subtle nonclassical feature of quantum mechanics.
That is, one and the same physical observable may
appear different, depending on the {\em context} of measurement;
i.e., depending on the particular way it was inferred. Stated
differently, the outcome of a physical measurement
may depend also on other physical measurements
which are coperformed.
In Bell's own words \cite[section 5]{bell-66}, ``The result of an
observation may
reasonably depend not only on the state of the system $\ldots $ but also
on the complete disposition of the apparatus.'' This property is usually
referred to as contextuality.

Formally, contextuality may be related to the nonexistence of two-valued
measures on Hilbert logics
\cite{birkhoff-36,kamber65,ZirlSchl-65,kochen1,svozil-tkadlec,svozil-ql} and
the partial algebra of projection operators \cite{kochen2,kochen3} when
the dimension of the Hilbert space is higher than two.
Contextuality then expresses the impossibility to
construct consistently truth values of the whole physical system
by any arrangement of truth values of ``proper parts'' thereof.
The term ``proper
part'' refers to any maximal number of independent comeasurable observables
corresponding to commuting self-adjoint operators. In quantum logics
\cite{kalmbach-83,pulmannova-91}, these are denoted by boolean
subalgebras or
``blocks'' which can be represented by a single ``maximal'' observable.
The entirety of
all proper parts is then identified with the whole physical system. By
definition, no union of different proper parts can itself be a proper
part, because there are always observables in the constituents which
are not comeasurable with another observable from any other different
proper part. This does not exclude that, for Hilbert spaces of dimension
greater than two,
there may exist one or more elements of different proper parts which
coincide.
Indeed, we shall encounter a system with three observables $A,B$ and $C$
such that $[A,B]=[A,C]=0$, whereas $[B,C]\neq 0$.
Therefore, although the proper parts are
``classical
mini-universes'' by the way they are constructed, their whole is not,
because it presupposes
counterfactual reasoning. (Indeed, Specker
\cite[in German]{specker-60}
has been motivated by
scholastic speculations of the so-called ``infuturabilities,'' or
``possibilities'').

In what follows, we propose to test contextuality by an EPR-type
measurement of one and the same observable, but with different
comeasurable observables. The difference to the usual EPR-type setup is
the identity of the observables and the specific attention paid to
other comeasurable observables, which are usually disregarded.
(For similar considerations, see an article by Heywood and Redhead \cite{hey-red}.)
Any argument of this kind must necessarily involve Hilbert spaces of
dimension higher than two, since because of orthogonality, for two or
lower dimensions,
if two observables are identical, all other comeasurable observables are
identical as well.

We shall adopt the original system of observables used by Kochen and
Specker
\cite{kochen1}. These observables are defined in threedimensional
Hilbert space and are based upon the spin one
observable  along an arbitrary unit vector
$(x_1,x_2,x_3)=(\sin \theta \cos \phi ,\sin \theta \sin \phi ,\cos \theta )\in
{\Bbb R}^3$, with polar coordinates $0\le \theta \le \pi$ and
$0\le \phi < 2\pi$. The radius is set to unity.
The corresponding hermitian
$(3\times
3)$-matrix  is given by
\begin{equation}
\label{l-soksp}
S(\theta , \phi )=
\left(
\begin{array}{ccc}
\cos \theta & {e^{-i\phi}\sin \theta \over \sqrt{2}}& 0      \\
{e^{i\phi}\sin \theta \over \sqrt{2}}& 0
& {e^{-i\phi}\sin \theta \over \sqrt{2}}      \\
0& {e^{i\phi}\sin \theta \over \sqrt{2}}& -\cos \theta
\end{array}\right).
\end{equation}

The spin
state observables
$J_1,J_2,J_3$
($\hbar =1$)
along the three cartesian coordinate axes $x=(1,0,0)\equiv ({\pi /
2},0,1)$, $y=(0,1,0)\equiv ({\pi / 2},{\pi
/ 2},1)$ and $z=(0,0,1) \equiv (0,0,1)$ are just given by
 (cf. Gudder
\cite[pp. 54-57]{gudder})
\begin{equation}
J_1= S({\pi \over 2},0),
\;
J_2= S({\pi \over 2},{\pi \over 2}),
\;
J_3= S(0,0),
\label{l-j123}
\end{equation}
and
\begin{equation}
\label{ansatz-so}
S^\ast (x_1,x_2,x_3)=
x_1J_1+
x_2J_2+
x_3J_3 ,
\end{equation}
where the asterisk ``$\ast$'' indicates that the arguments are
the usual cartesian coordinates.
Spin state measurements along another orthogonal tripod
$\bar{x},\bar{y},\bar{z}$ can be
easily represented by $S(\bar{x}),S(\bar{y})$ and $S(\bar{z})$.

Consider now
the squares of the spin state observables introduced in Equation
(\ref{l-j123}).
\begin{equation}
J_1^2= {1\over 2}
\left(
\begin{array}{ccc}
1&0&1\\
0&2&0\\
1&0&1
\end{array}
\right), \;
J_2^2= {1\over 2}
\left(
\begin{array}{ccc}
1&0&-1\\
0&2&0\\
-1&0&1
\end{array}
\right), \;
J_3^2=
\left(
\begin{array}{ccc}
1&0&0\\
0&0&0\\
0&0&1\\
\end{array}
\right).
\end{equation}

Let us consider another system of $\bar{J}_i^2$'s rotated by $\phi\neq 0
\textrm{ mod } \pi /2$
along the
$z=(0,0,1)$-axis. According to
Equation
(\ref{l-soksp}),
\begin{equation}
{\bar J}_1^2= \left(S({\pi \over 2}, {\phi })\right)^2, \quad
{\bar J}_2^2= \left(S({\pi \over 2}, \phi +{\pi \over 2})\right)^2,
\quad
{\bar J}_3^2=\left(S(0,0)\right)^2.
\end{equation}

By inspection, it can be verified that the $J_i^2$'s and $\bar{J}_i^2$'s
form two mutually commuting systems; i.e.,
\begin{eqnarray*}
&[J_1^2,J_2^2]=[J_1^2,J_3^2]=[J_2^2,J_3^2]=0,&\\
&[\bar{J}_1^2,\bar{J}_2^2]=[\bar{J}_1^2,\bar{J}_3^2]=[\bar{J}_2^2,\bar{J}_3^2]=0.&\\
\end{eqnarray*}
But not all $J_i^2$'s commute with all $\bar{J}_i^2$'s. For instance,
$[\bar{J}_1^2,J_1^2]\neq 0.$
Indeed, only
$J_3^2$ commutes with $\bar{J}_3^2$, because these operators are
identical.

As has already been pointed out by von Neumann
\cite{v-neumann-49,halmos-vs,kochen2,riesz-nagy}, for any system of
mutually commuting self-adjoint operators $H_1,H_2,\ldots$ there exists
a maximal operator $U$ such that all $H_i$'s are functions $f_i$ of $U$;
i.e.,
\begin{equation}
H_i=f_i(U).
\end{equation}

Applying this result to the two systems of mutually commuting operators
$\{
J_1^2,
J_2^2,
J_3^2\}$
and
$\{
\bar{J}_1^2,
\bar{J}_2^2,
\bar{J}_3^2\}$
yields two maximal operators $U$ and $\bar{U}$ and sets of functions $f_i$,
$\bar{f}_i$ such that
\begin{equation}
J_i^2=f_i(U) \quad {\rm and }\quad
\bar{J}_i^2=\bar{f}_i(\bar{U}), \quad i=1,2,3.
\end{equation}
In particular,
\begin{equation}
J_3^2=f_3(U)=\bar{f}_3(\bar{U})=\bar{J}_3^2.
\end{equation}

More explicitly \cite{kochen2,svozil-ql}, let $a\neq b\neq c\neq a$ and
\begin{eqnarray}
U&=&aJ_1^2+bJ_2^2+cJ_3^2=
 {1\over 2}
\left(
\begin{array}{ccc}
a+b+2c&0&a-b\\
0&2a+2b&0\\
a-b&0&a+b+2c
\end{array}
\right),
\label{le-ndfuo}\\
\bar U (\phi)&=& \bar a\bar J_1^2+\bar b\bar J_2^2+\bar c\bar J_3^2=
 {1\over 2}
\left(
\begin{array}{ccc}
\bar a + \bar b + 2 \bar c      & 0        & (\bar a - \bar b)e^{-2i\phi}\\
0                & 2\bar a + 2\bar b  & 0 \\
(\bar a - \bar b)e^{2i\phi} & 0        &\bar a + \bar b + 2 \bar c
\end{array}
\right).
\end{eqnarray}
The diagonal form of $U$ and $\bar U$ is
${\rm diag}(a+b,b+c,a+c)$
and
${\rm diag}(\bar a+\bar b,\bar b+\bar c,\bar a+\bar c)$
respectively.
Measurement of $U$  can, for instance, be realized by a set
of beam splitters
\cite{rzbb}; or in an arrangement proposed by Kochen and Specker
\cite{kochen1}.
Any such measurement will yield either the eigenvalue
$a+b$  (exclusive) or the eigenvalue
$b+c$  (exclusive) or the eigenvalue
$a+c$.
Since $a,b,c$ are mutually distinct, the eigenvalues
of $U$ are nondegenerate.

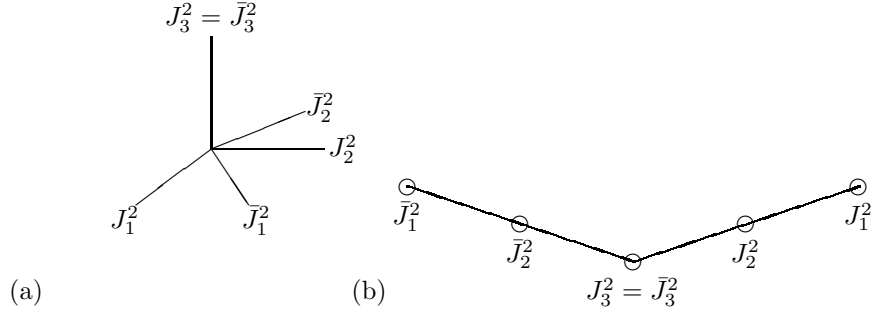
\begin{figure}
\begin{center}
(a)
\unitlength 0.50mm
\linethickness{0.4pt}
\begin{picture}(75.00,75.00)
\put(40.00,40.00){\line(0,1){30.00}}
\put(40.00,40.00){\line(1,0){30.00}}
\put(40.00,40.00){\line(-4,-3){20.00}}
\put(40.00,40.00){\line(2,-3){10.00}}
\put(40.00,40.00){\line(5,2){25.00}}
\put(40.00,75.00){\makebox(0,0)[cc]{$J_3^2=\bar J_3^2$}}
\put(17.00,21.00){\makebox(0,0)[cc]{$J_1^2$}}
\put(75.00,40.00){\makebox(0,0)[cc]{$J_2^2$}}
\put(52.00,21.00){\makebox(0,0)[cc]{$\bar J_1^2$}}
\put(69.00,51.00){\makebox(0,0)[cc]{$\bar J_2^2$}}
\end{picture}
(b)
\unitlength 1.00mm
\linethickness{0.4pt}
\begin{picture}(61.33,16.00)
\multiput(0.33,15.00)(0.36,-0.12){84}{\line(1,0){0.36}}
\multiput(30.33,5.00)(0.36,0.12){84}{\line(1,0){0.36}}
\put(30.33,5.00){\circle{2.00}}
\put(45.33,10.00){\circle{2.00}}
\put(60.33,15.00){\circle{2.00}}
\put(0.33,15.00){\circle{2.00}}
\put(15.33,10.00){\circle{2.00}}
\put(60.33,11.00){\makebox(0,0)[cc]{$J_1^2$}}
\put(45.33,6.00){\makebox(0,0)[cc]{$J_2^2$}}
\put(30.33,1.00){\makebox(0,0)[cc]{$J_3^2=\bar J_3^2$}}
\put(15.33,6.00){\makebox(0,0)[cc]{$\bar J_2^2$}}
\put(0.33,11.00){\makebox(0,0)[cc]{$\bar J_1^2$}}
\end{picture}
\end{center}
\caption{\label{f-ffiab2}(a) Two orthogonal tripods with a common leg $J_3^2=\bar J_3^2$.
 (b) The same configuration represented by the
associated Greechie diagram.  }
\end{figure}

At the same time, $J_1^2,J_2^2,J_3^2$
are orthogonal projection operators in
${\Bbb R}^3$: they are idempotent $J_i^2J_i^2=J_i^2$, with eigenvalues
$0$ and $1$ for
$i=1,2,3$.  (The same is true for any system $\bar J_1^2,\bar
J_2^2,\bar J_3^2$.) Alternatively, they can be identified with
onedimensional orthogonal subspaces of ${\Bbb R}^3$ which in turn are
spanned by the orthogonal vectors $(1,0,0)$, $(0,1,0)$ and $(0,0,1)$.
In quantum logic, they can be identified with atomic
propositions and can be conveniently represented by
hypergraphs called
``Greechie diagrams,'' in which points represent atoms and all
orthogonal
atoms belonging to the same tripod are represented by edges or smooth
curves. The spatial configuration of subspaces as well as the associated
Greechie diagram of the combined systems
$J_1^2,J_2^2,J_3^2$ and $\bar J_1^2,\bar J_2^2,\bar J_3^2$ are drawn in
Figure \ref{f-ffiab2}.

The $J_3^2$ and $\bar J_3^2$ are then polynomials of $U$ and $\bar U$, respectively; i.e.,
\begin{eqnarray}
J_3^2 =f_3(U)&=&
 {1\over (c-b)(a-c)}\left[ U^2-U(a+b+2c)+2(a+b)c{\Bbb I}\right],\\
\bar J_3^2 =\bar f_3(\bar U)&=&
 {1\over (\bar c-\bar b)(\bar a-\bar c)}\left[ \bar U^2-\bar U(\bar
a+\bar b+2\bar c)+2(\bar a
+\bar b)\bar c{\Bbb I}\right].
\end{eqnarray}
Furthermore,
\begin{equation}
J_1^2+
J_2^2+
J_3^2=
\bar{J}_1^2+
\bar{J}_2^2+
\bar{J}_3^2 =2{\Bbb I},
\end{equation}
indicating that since the possible eigenvalues of any $J_i^2,
i=1,2,3$ are
either~0 or~1, the eigenvalues of two observables $J_i^2,i=1,2,3$ must
be~1, and one must be~0.
Thus any measurement of the
maximal operator $U$ yields
$a+b$ associated with $J_1^2=J_2^2=1$, $J_3^2=0$ (exclusive) or
$a+c$ associated with $J_1^2=J_3^2=1$, $J_2^2=0$ (exclusive) or
$b+c$ associated with $J_2^2=J_3^2=1$, $J_1^2=0$.

Now consider an EPR-type arrangement with two particles in an identical state
\begin{eqnarray}
{1\over \sqrt{3}}&&
\left[
\vert a+b\rangle \vert a+b\rangle
+
\vert a+c\rangle \vert a+c\rangle
+
\vert b+c\rangle \vert b+c\rangle
\right]
\label{eq-coun-in}
\\
&&
\equiv
{1\over \sqrt{3}}
\left[
\left(\begin{array}{c} 0\\1\\0\end{array}\right)
\times
\left(\begin{array}{c} 0\\1\\0\end{array}\right)
+
\left(\begin{array}{c} 1\\0\\1\end{array}\right)
\times
\left(\begin{array}{c} 1\\0\\1\end{array}\right)
+
\left(\begin{array}{c} -1\\0\\1\end{array}\right)
\times
\left(\begin{array}{c} -1\\0\\1\end{array}\right)
\right]
\nonumber
\end{eqnarray}
The quantum numbers $a+b$, $a+c$ and $b+c$ refer to
the eigenstates of $U$ with eigenvalues $a+b$, $a+c$ and $b+c$, respectively.
The eigenvalues and eigenstates  of $\bar U$  are $\bar a+\bar b$, $\bar a+\bar c$, $\bar b+\bar c$ and
$(0,1,0)$, $(e^{-2i\phi},0,1)$, and $(-e^{-2i\phi},0,1)$, respectively.

Now consider the following question: assume $U$
and
$\bar U$ are measured for the right and the left particle separately.
(Of course, one may also successively measure
$\{
J_1^2,
J_2^2,
J_3^2\}$
and
$\{
\bar{J}_1^2,
\bar{J}_2^2,
\bar{J}_3^2\}$.)
{\em Would the outcome of the measurement of $J_3^2=\bar J_3^2$ be
different, depending on whether it was derived from $U$ or $\bar U$?}

There are at least three alternative answers which will be discussed
shortly.

{\em ``Strong'' contextuality assumption (I):
Although
$J_3^2$ and
$\bar J^2_3$ are identical operators,  $U$ and $\bar U$ are not;
and
the associated measurement results need {\em not} coincide}, since
$J_3^2$ has to be inferred from $U$ (or equivalently, comeasured with
$J_1^2$  and
$J_2^2$),
while
$\bar J_3^2$
has to be be inferred from $\bar U$ (or equivalently, comeasured with
$\bar J_1^2$ and
$\bar J_2^2$). This would then make it necessary to add to each quantum
number also---in Bell's terms---the complete disposition of the
measurement apparatus, which can be represented by the associated
maximal operator.
%

We may quantify ``strong'' contextuality by
noticing  that the $J^2_i$'s are dichotomic observables with
eigenvalues $0$ and $1$. Therefore, in analogy to EPR-type experiments,
it is possible to define a correlation function
\begin{equation}
C(\phi )=\lim_{N\rightarrow \infty}{1\over N}\sum_{j=i}^N
r_j(J_3^2)r_j(\bar J_3^2),
\label{e-noncor}
\end{equation}
where $N$ is the number of experiments, the index $j$ denotes the $j$'th
experiment; and $r(1)=+1$ and $r(0)=-1$. $\phi $ is the relative
angle between the $x$- and $\bar x$-axes. If both axes coincide, then
$C(0)=1$.

The proposed experiment tests ``strong'' contextuality in the following way. If
$C(
\phi
)=1$ for
$0< \phi <\pi /2$, then the system behaves noncontextually.
This can be  verified by considering formula
(\ref{e-noncor}): in the noncontextual case,
$r_j(J_3^2)$ and $r_j(\bar J_3^2)$ are always the same ($+1,+1$ or $-1,-1$),
hence their product always yields $1$.
Contextuality manifests itself as $C(\phi)<1$ for some value of
$\phi$.
In this case,
$r_j(J_3^2)$ and $r_j(\bar J_3^2)$ differ sometimes ($+1,-1$ or $-1,+1$),
resulting in a negative product which reduces the overall sum in
(\ref{e-noncor}).

In view of the highly counterintuitive consequences discussed for the
contextual case, let us consider
{\em noncontextuality assumption (II):
$J_3^2$ and
$\bar J^2_3$ are identical operators which therefore {\em always yield
identical} observations, independent of the way they have been derived.}
As innocent and evident this statement may appear, it clashes with a
theorem derived by
Kamber \cite{kamber65}, Zierler and Schlessinger \cite{ZirlSchl-65} and
Kochen and Specker
\cite{kochen1}. (For recent reviews, see
\cite{redhead,peres,mermin-93,svozil-tkadlec,svozil-ql}, among others.)
For then we could consider instead of a two-particle EPR-experiment
a, say, 17-entangled particle experiment characterized by a similar
state as in
(\ref{eq-coun-in})
and make measurements of $U$ (and thus of  $J_1^2,J_2^3$ and $J_3^2$) along  the 17
direction vectors $(0,0,1)$,  $(0,1,0)$
and all coordinate permutations from $(0,1,\sqrt2)$, $(1,\pm1,\sqrt2)$. This system
has been
suggested by Peres \cite{peres-91,peres}, but the original Kochen-Specker configuration
or any other proper system of vectors would do just as well (for a review, see,
e.g., \cite{mermin-93,svozil-tkadlec,svozil-ql}).
In this case, the assumption of noncontextuality results in a complete
contradiction with the noncontextuality assumption (II):
after measuring all 17 particles and checking the appropriate
observables it turns out that the outcome of measurements of at least
two observables which
correspond to identical operators but are measured alongside different
coobservables (blocks) are different.

There is yet another, more subtle alternative which will be called
{\em ``haunted'' contextuality assumption (III): }
contextuality never manifests itself in its ``strong'' form (I) but only through counterfactual
reasoning. Insofar states of the form (\ref{eq-coun-in}) are explicitly
constructed to yield identical results on measurements of $J_3^2$,
these measurement outcomes are independent of the way they have been
inferred;  and in particular what observables have been measured
alongside. This does not contradict the Kochen-Specker theorem, since
obviously states obeying such perfect correlations can be constructed
only in {\em one} direction, wheres the Kochen-Specker theorem necessarily requires
directions associated with noncomeasurable, complementary observables.
Thus the test of ``strong'' contextuality
will fail; i.e., $C(\phi )$ will always be unity. But this does not
exclude---and indeed makes necessary---arguments involving
counterfactuality, such as the Kochen-Specker theorem or the
GHZ-theorem which, although of doubtless importance conceptually,
bear a non-operational flavor
in the sense of direct physical testability.



\end{document}